\documentclass[a4paper,10pt,oneside,twocolumn]{article}
\usepackage{graphicx}
\usepackage{mathptmx}
\usepackage{titlesec}
\titlespacing*{\section}
{0pt}{4.3ex minus .2ex}{0pt}
\titlespacing*{\subsection}
{0pt}{4.3ex minus .2ex}{0pt}
\usepackage{titling}
\pretitle{\vspace{-4cm}
\textit{
\\
\\
\begin{small}
Accepted for presentation at the Audio Engineering Society (AES) 156th Convention, June 2024, Madrid, Spain.
\end{small}
\\
\\
}

\par\noindent\rule{\textwidth}{1pt}
\newline\vspace{1em}\newline
\huge\bfseries}
\posttitle{\vspace{1em}}
\preauthor{\newline}
\postauthor{\newline}
\date{}

\usepackage{authblk}

\usepackage[hyphens]{url}
\usepackage[hidelinks]{hyperref}

\usepackage{cleveref} % 
\crefname{figure}{Figure}{Figures}
\Crefname{figure}{Figure}{Figures}
\crefname{appendix}{Appendix}{Appendices}
\Crefname{appendix}{Appendix}{Appendices}
\crefname{equation}{Equation}{Equations} 
\Crefname{equation}{Equation}{Equations}
\crefname{table}{Table}{Tables} 
\Crefname{table}{Table}{Tables}
\crefname{chapter}{Chapter}{Chapters} 
\Crefname{chapter}{Chapter}{Chapters}
\crefname{subsection}{Section}{Subsection} 
\Crefname{subsection}{Section}{Subsection}
\crefname{section}{Section}{Section} 
\Crefname{section}{Section}{Section}

\usepackage[utf8]{inputenc}
\usepackage{microtype}
\usepackage[numbers,square]{natbib}
\usepackage{booktabs}
\usepackage{xcolor}
\usepackage{array}
\usepackage{enumitem}
\usepackage{multirow}
\usepackage{balance}
\usepackage[nolist,nohyperlinks]{acronym}

\newcommand\blfootnote[1]{%
  \begingroup
  \renewcommand\thefootnote{}\footnote{#1}%
  \addtocounter{footnote}{-1}%
  \endgroup
}

\begin{acronym}
\acro{OBA}{Object-Based Audio}
\acro{SLD}{Speech Loudness Deviation}
\acro{SBLD}{Speech-to-Background Loudness Difference}
\acro{QC}{quality control}
\acro{SNR}{Signal-to-Noise Ratio}
\acro{SRT}{Speech Reception Threshold}
\acro{RMS}{Root Mean Square}
\acro{MAE}{Mean Absolute Error}
\acro{DNN}{Deep Neural Network}
\acro{ME}[M\&E]{Music and Effects}
\acro{ESTOI}{Extended Short-Time Objective Intelligibility }

\acro{LDR}{Loudness-to-Dialog Ratio}
\acro{LRA}{Loudness Range}

\acro{SL}{clean Speech Loudness}
\acro{BL}{Background Loudness}
\acro{ST-SL}{Short-term clean Speech Loudness}
\acro{ST-BL}{Short-term Background Loudness}
\acro{ST-SBLD}{Short-term Speech-to-Background Loudness Difference}

\acro{DPP}{Digital Production Partnership}
\acro{AES}{Audio Engineering Society}
\acro{EBU}{Europea Broadcasting Union}

\acro{VAD}{Voice Activity Detection}
\end{acronym}

\title{Speech Loudness in Broadcasting and Streaming}
\author[1]{Matteo Torcoli}
\author[1]{Mhd Modar Halimeh}
\author[1]{Thomas Leitz}
\author[1]{Yannik Grewe}
\author[1]{Michael Kratschmer}
\author[2]{Bernhard Neugebauer}
\author[1]{Adrian Murtaza}
\author[1]{Harald Fuchs}
\author[3]{Emanuël A. P. Habets}

\affil[1]{Fraunhofer Institute for Integrated Circuits IIS, Erlangen, Germany}
\affil[2]{DSP Solutions, Regensburg, Germany}
\affil[3]{International Audio Laboratories Erlangen\textsuperscript{$\ast$}, Erlangen, Germany}

\begin{document}

\twocolumn[
\maketitle % MANDATORY!

\begin{abstract}
The introduction and regulation of loudness in broadcasting and streaming brought clear benefits to the audience, e.g., a level of uniformity across programs and channels.
Yet, speech loudness is frequently reported as being too low in certain passages, which can hinder the full understanding and enjoyment of movies and TV programs.
This paper proposes expanding the set of loudness-based measures typically used in the industry.
We focus on speech loudness, and we show that, when clean speech is not available, \acp{DNN} can be used to isolate the speech signal and so to accurately estimate speech loudness, providing a more precise estimate compared to speech-gated loudness.
Moreover, we define critical passages, i.e., passages in which speech is likely to be hard to understand.
Critical passages are defined based on the local \ac{SLD}
and the local \ac{SBLD}, as \ac{SLD} and \ac{SBLD} significantly contribute to intelligibility and listening effort.
In contrast to other more comprehensive measures of intelligibility and listening effort, \ac{SLD} and \ac{SBLD} can be straightforwardly measured, are intuitive, and, most importantly, can be easily controlled by adjusting the speech level in the mix or by enabling personalization at the user's end.
Finally, examples are provided that show how the detection of critical passages can support the evaluation and control of the speech signal during and after content production. 
\vspace{1em}
\end{abstract}
] % and of two columns

\acresetall

\section{Introduction}

\subsection{Problem Formulation\\and Contributions}

Speech in broadcast and streamed content is frequently reported as difficult to understand~\cite{torcoli2021dialog, mapp2016intelligibility}.\blfootnote{\textsuperscript{$\ast$} A joint institution of Fraunhofer IIS and Friedrich-Alexander-Universit{\"a}t Erlangen-N{\"u}rnberg (FAU), Germany.}
One of the main causes is the low level with which speech is mixed in the soundtrack~\cite{mathers1991study, thornton:loudness}.
Some unintelligible speech can be part of the original artistic intent. However, speech is meant to be understood in the vast majority of cases.

Nowadays, there exist solutions allowing the audience to personalize the audio mix and adapt it to their needs, devices, and preferences~\cite{torcoli2021dialog, ward2019casualty, rieger2023dialogue, amazon2023}. 
There are also recommendations and guidelines focusing on speech levels, e.g., in cinematic content~\cite{ebu_s4}. 

As problems with speech intelligibility and listening effort continue to be reported, this paper aims to contribute to establishing procedures and tools that effectively assist the production and quality control of speech within soundtracks.
To do so, an assistive framework is proposed, which includes novel measures based on \mbox{ITU-R} BS.1770 loudness~\cite{itu_bs1770}. These can complement the set of loudness-based measures that are already commonly used in the industry. 

The key is to accurately measure speech loudness over time, analyze its dynamics, and its local relation to other background sounds such as \ac{ME}. Based on this analysis, critical speech passages are defined, i.e., passages in which speech is likely to be hard to understand. Being simply based on local loudness measures, this analysis does not cover all possible cases of hard-to-understand speech, e.g., unfamiliar accent or vocabulary. However, it has the advantage of being simple and predictable, and offers a straightforward way to resolve the detected critical passages by changing the speech level in the mix, or by providing the audience with personalization features.

Further contributions of this paper are: 1) Reviewing the concepts of intelligibility and listening effort in the context of audio for broadcasting and streaming, and how they relate to ITU-R BS.1770; 2) Evaluating the recent progress in deep learning to estimate speech loudness when the clean speech signal is not available.

\subsection{Data Description}
\label{sec:data}
A number of experimental results are reported throughout this paper.
These experiments are carried out using an internal dataset, carefully constructed to represent the large diversity of audio signal types encountered in broadcasting and streaming.
The dataset contains matching pairs of stereo audio signals sampled at $48$~kHz.
Each pair consists of a clean speech signal and a background signal containing \ac{ME}.
Speech signals comprise different languages, male and female speakers, both adults and children. Different speaking styles and emotions are included, e.g., shouting and whispering. 
A large variation of background elements is represented, e.g., music, cinematic effects, and environmental noises such as stadium, babble noise, traffic, and outside nature.
Short broadcast-like audio mixes are obtained by superimposing these pairs of signals.
Some of these mixes constitute excerpts taken from real-world broadcast TV programs, while others were produced internally for demo purposes.
The total duration of the mixed pairs is $1$ hour, and the average duration is $10$ s. 
For illustrative purposes, the original mixing ratio between speech and \ac{ME} is not used. Instead, the mixing ratio is artificially controlled, or only clean speech is used, as detailed in each experiment. In any case, the full $1$ hour is used at each mixing ratio.

The focus of this dataset is on the simultaneous presence of speech and \ac{ME}. Real-world movies and TV programs can feature numerous parts where speech is present with little to no \ac{ME}.
Considering speech-only parts would make it easier to estimate some of the measures reported in the following, and so average errors could be smaller in practice. Still, it is important to focus on mixtures where speech and \ac{ME} are simultaneously active to better understand the most critical scenario, i.e., the one dominating estimation errors in the final applications. The same reasoning applies to the focus on short excerpts instead of full programs, with the addition that a dataset of full programs with separate clean speech is not available.

\subsection{Paper Structure}

The paper is structured as follows. \cref{sec:intel} reviews intelligibility and listening effort. \cref{sec:loud} gives an overview of ITU-R BS.1770 loudness. \cref{sec:sLoud} considers speech loudness as a proxy for intelligibility and listening effort, and highlights the differences between speech-gated loudness and true speech loudness. \cref{sec:proposed} details the proposed framework and evaluates the performance of a speech separation technique used to estimate speech loudness and its relation to \ac{ME}. The employed speech separation system utilizes deep learning and reflects the latest advancements in this field.

\section{Understanding Speech}
\label{sec:intel}
Intelligibility and listening effort are among the main attributes characterizing the perception of speech. 
Intelligibility assesses whether \textit{what} is said can be identified.
Listening effort describes \textit{how easy/difficult} it is to identify and understand what is said.
Additional speech attributes can refer to \textit{how} speech sounds (e.g., timbre).

\subsection{Intelligibility}

Intelligibility is quantified by counting the number of words or phonemes correctly identified.
When speech is presented alone without any interfering source, the absolute level of speech is one of the main factors directly determining the intelligibility level \cite{loizou2007}.
In addition, the \ac{SNR} affects intelligibility when speech is presented together with interfering sounds \cite{ohlenforst2018impact}. In our case, \textit{noise} in \ac{SNR} refers to all audio sources that are not speech, e.g., \ac{ME}.

\begin{figure}[t]
\begin{center}
\includegraphics[width=1\columnwidth]{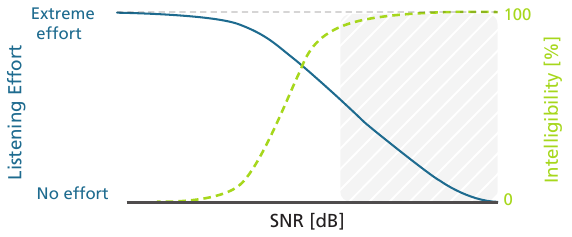}
\caption{Illustration of typical relation between speech intelligibility (dashed line) and listening effort (solid line) with respect to SNR \cite{ohlenforst2018impact, klink2012measuring}. Significant effort can be experienced even if intelligibility is perfect or close to perfect (grey area).}
\label{fig:SNR}
\end{center}
\end{figure}

\begin{figure}[t]
\begin{center}
\includegraphics[width=1\columnwidth]{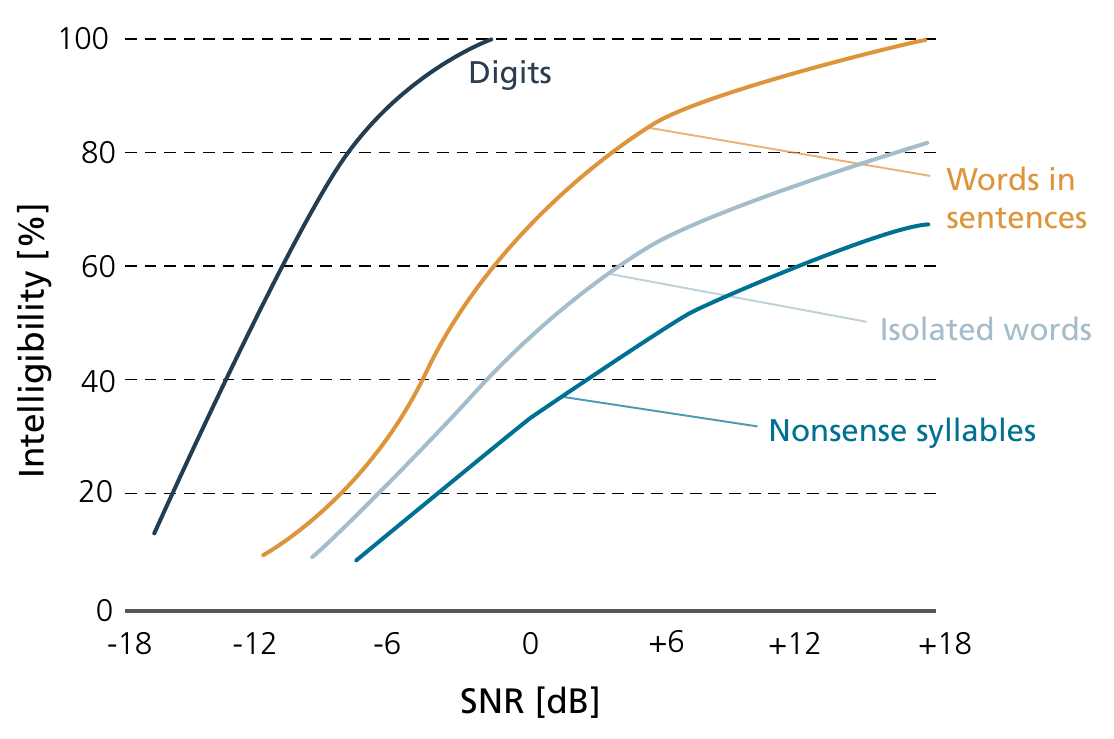}
\caption{Effect of SNR and nature of the signal on its intelligibility. Intelligibility increases with \ac{SNR} as well as with size and type of vocabulary and semantic context of the signal \cite{rood2006flight}.}
\label{fig:SNR_b}
\end{center}
\end{figure}

Following~\cite{loizou2007, klink2012measuring}, the typical sigmoid-shaped trend of intelligibility as a function of \ac{SNR} is shown in \cref{fig:SNR}.
The \ac{SNR} level at which listeners identify words with 50\% accuracy is referred to as the \ac{SRT}. For instance, \ac{SRT} was measured to be about 3\,dB for standardized sentences presented with speech-shaped noise~\cite{nilsson1994}. The nature and the context of the signal can change the \ac{SRT} significantly, as shown in \cref{fig:SNR_b}. In any case, the \ac{SNR} has a direct effect on intelligibility. 
It has to be noted that in these audiology studies, signal intensities (and \ac{SNR}) are computed based on the ubiquitous \ac{RMS} energy level of the signals, expressed in decibels (dB).

\subsection{Listening Effort}

Intelligibility does not capture the effort required to recognize phonemes, words, or sentences. Perfect intelligibility can be observed along with a significant amount of effort, as depicted in \cref{fig:SNR}.
Listening effort can be defined as the cognitive resources (or cognitive load) required to perform a listening task~\cite{alhanbali2019measures}, e.g., understanding the plot while watching a movie.
The amount of cognitive resources allocated is not trivially measured, making listening effort often hard to quantify. Typical measures include self-report, behavioral performance, or physiological measures.
However, the scientific literature on listening effort is as ambiguous as it is voluminous~\cite{alhanbali2019measures}.
We propose to take a practical approach to this problem, building on the fact that absolute and relative levels of speech are among the main factors contributing to listening effort (\cref{fig:SNR}), as also shown in \cite{torcoli2022dialogue} with particular focus on broadcast audio material.

% Definitions for big table
\def\firstColW{4cm}
\def\secondColW{4.8cm}
\def\leftMarginItem{4mm}

\begin{table*}
 \caption{In broadcasting and streaming, many factors influence the final speech intelligibility and listening effort. This paper focuses on the factors influencing speech level at the production stage (\textcolor{orange}{\textbf{in orange}}).}
 \begin{footnotesize}
  \centering
  \setlength{\extrarowheight}{0.5em}
  \begin{tabular}{lll}
    \toprule
        \textbf{Stage} & \textbf{Influencing perceived speech level} & \textbf{Other relevant factors} \\ 
    
    \midrule

        \begin{tabular}{c}
            \textbf{Acting, Recording, Production}\\ %\midrule
    
            \parbox{\firstColW}{
            \vspace{3mm}
            \centering
            \includegraphics[width=2.5cm]{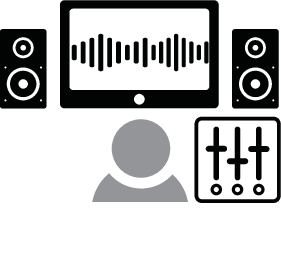}
            }
        \end{tabular}
    &

        \parbox{\secondColW}{
        \begin{itemize}[leftmargin=\leftMarginItem]
            \item \textcolor{orange}{\textbf{Program loudness}}
            \item \textcolor{orange}{\textbf{Program dynamic range}}
            \item \textcolor{orange}{\textbf{Absolute speech loudness}}
            \item \textcolor{orange}{\textbf{Relative speech loudness}}
            \item \textcolor{orange}{\textbf{Speech dynamic range}}
        \end{itemize}}
    &
        \parbox{\secondColW}{
        \begin{itemize}[leftmargin=\leftMarginItem]
            \item Articulation, clarity of speech, and speech rate
            \item Sentence complexity
            \item Speaker Accent            
            \item Accompanying visual and acoustic cues
        \end{itemize}}
    \vspace{-0.5cm}
    \\ 
    \midrule

        \begin{tabular}{c}
            \textbf{Receiving and Playback}\\ %\midrule
            \parbox{\firstColW}{
            \vspace{3mm}
            \centering
            \includegraphics[width=2.0cm]{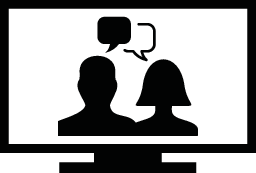}
            }
        \end{tabular}
    &
        \parbox{\secondColW}{
        \begin{itemize}[leftmargin=\leftMarginItem]
            \item Level of audio playback and device settings
            \item Type and quality of the device
            \item On-device processing
            \item Selected object-based preset
        \end{itemize}}

    &
        \parbox{\secondColW}{
        \begin{itemize}[leftmargin=\leftMarginItem]
            \item Audio compression
            \item Bandwidth and bitrate
            \item Packet loss
            \item Non-linear distortions
            
        \end{itemize}}
    \\ \midrule

        \begin{tabular}{c}
            \textbf{Listener and Environment}\\ %\midrule
            \parbox{\firstColW}{
            \vspace{3mm}
            \centering
            \includegraphics[width=2.2cm]{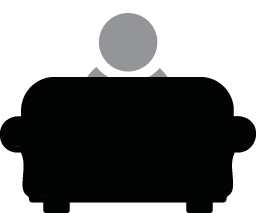}
            }
        \end{tabular}
    &
        \parbox{\secondColW}{
        \begin{itemize}[leftmargin=\leftMarginItem]
            \item Hearing loss
            \item Environmental noise level         
            \item If on loudspeakers: Setup and calibration, room reverberation, distance between listener and loudspeakers
            \item If with wearables or hearing aids: Additional processing (noise canceling).
        \end{itemize}}

    &
        \parbox{\secondColW}{
        \begin{itemize}[leftmargin=\leftMarginItem]
            \item Personal taste
            \item Familiarity with language, dialect, accent, or topics
            \item Attention and fatigue
        \end{itemize}}
    \\    
    \bottomrule
  \end{tabular}
  \label{tab:stages}
\end{footnotesize}
\end{table*}

\subsection{Speech Level in Broadcasting\\and Streaming}
\label{sec:speech}
We already highlighted the importance of absolute and relative speech levels in determining intelligibility and listening effort.
In turn, the speech level perceived while consuming audiovisual content is determined by many factors encountered across different stages, as summarized in \cref{tab:stages}.

In this work, we focus on the factors influencing speech level at the production stage.
At this stage, loudness-based measures are commonly used.

\section{ITU-R BS.1770 Loudness}
\label{sec:loud}
\subsection{Overview}
Loudness as per ITU-R BS.1770 \cite{itu_bs1770} is an objective measure with the goal of matching the subjective loudness impression of an acoustic signal. It is measured in LKFS (Loudness, K-weighted, relative to Full Scale) or equivalently in Loudness Unit relative to Full Scale (LUFS) \cite{ebu_r128}.
When relative to another loudness value, Loudness Unit (LU) is simply used.
LU is equivalent to dB in that a broadband level gain of $X$\,dB causes the loudness to change by $X$\,LU, if remaining within the defined range, e.g., $(-70, +3)$\,LUFS for stereo. This equivalence with the dB scale is not only a useful property, but it also allows for a direct link to the works from audiology studying the impact of absolute and relative speech levels (in dB) on intelligibility and listening effort (\cref{sec:intel}).

\begin{figure}[t]
\begin{center}
\includegraphics[width=0.80\columnwidth]{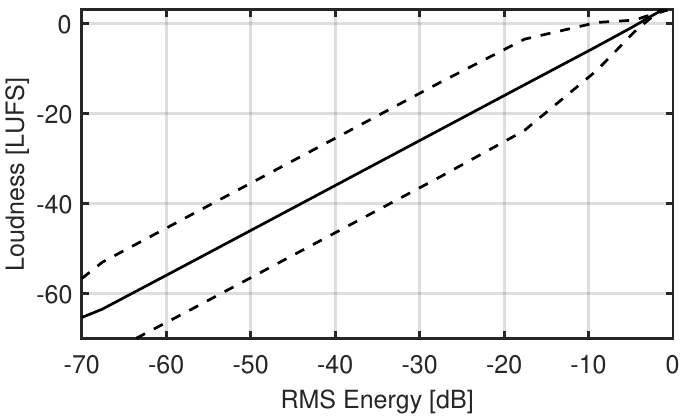}
\caption{Average (solid line) $\pm$ standard deviation (dashed lines) showing the relation between signal \acf{RMS} energy level and integrated loudness for stereo signals containing studio recordings of speech only.}
\label{fig:testa}
\end{center}
\end{figure}

At the core of ITU-R BS.1770, energy is computed as the mean square for each channel, tying loudness to \ac{RMS} and \ac{SNR}, as shown in \cref{fig:testa,fig:testa_b} (using the data described in \cref{sec:data}). 
Before the energy computation, ITU-R BS.1770 applies a frequency weighting to account for the frequency-dependent auditory sensitivity of humans. 
Hence, a weighted sum of the channels' energy values is performed.
Finally, a gating mechanism is specified, excluding some parts of the signal from the integration windows. This gating is based on an absolute threshold ($-70$ LUFS) and on a second threshold relative to the level measured after the application of the first threshold.
Other gating mechanisms are applied in the industry, e.g., based on speech activity within the signal, as discussed in \cref{sec:gating}.

Although not perfect \cite{berendes2022validating}, ITU-R BS.1770 became the standard for measuring audio levels in broadcasting and streaming. Nowadays, it contributes significantly to uniform audio levels across programs, channels, and platforms. It is considered the main tool with which regional recommendations such as \cite{ebu_r128, atsc_a85} were able to end the so-called loudness war~\cite{cramer2010} and the continuous need to adjust the playback volume on the user's side. 

\subsection{Integration Times}

Loudness can be measured considering different integration times. Three time-scales are defined~\cite{ebu_3341}:
\begin{enumerate}
    \item \textbf{Integrated} Loudness: The full audio program is considered from beginning to end.
    \item \textbf{Short-term} Loudness: A moving time window of 3\,s is considered.
    \item \textbf{Momentary} Loudness: A  moving time window of 400\,ms is considered.
\end{enumerate}

\subsection{Existing Recommendations}

Numerous recommendations exist providing target values for the loudness of broadcast and streamed programs\footnote{A more complete overview of the existing loudness recommendations is provided at: \url{https://www.rtw.com/en/blog/worldwide-loudness-delivery-standards.html} [Accessed Feb. 2024].}, e.g., \cite{ebu_r128, atsc_a85, aes77_2023}.
These recommendations are centered around normalizing content to a specific loudness level, i.e., by applying a time-independent broadband gain to the full soundtrack. Some dynamic aspects are also covered, e.g., by considering \ac{LRA} or maximum momentary or short-term loudness.
%
%These recommendations are valid in different regions of the world and for different types of content and service. In any case, 
The goal is to deliver a uniform experience to the audience with respect to program audio levels.
Parts of these recommendations focus on speech, as discussed in the following.

\section{Speech Loudness}
\label{sec:sLoud}

The importance of using speech loudness has been long known~\cite{riedmiller2005practical}, and speech loudness is often recommended instead of program loudness~\cite{atsc_a85, riedmiller2005practical, aes71_2018, netflixMixSpecs}.
The rationale is that speech is the element around which \ac{ME} are mixed, as well as the element based on which viewers perceive program loudness and set their playback level~\cite{netflixLRA}.
This section reviews the main loudness-based measures with a focus on speech, i.e., \ac{SL}, speech-gated loudness, \ac{LRA}, \ac{SBLD}, and \ac{LDR}.

\subsection{Speech Loudness and Speech Gating}
\label{sec:gating}

Ideally, \ac{SL} can be easily measured given the clean speech signal. However, clean speech alone is often not available.
The community is well aware of this problem, e.g., AES77~\cite{aes77_2023} considers both the case in which speech is available and the one in which it is not.
The common way to deal with the unavailability of clean speech is to use speech-gated loudness. This is often referred to as dialog-gated or anchor loudness~\cite{atsc_a85, apple:anchor}. With respect to ITU-R BS.1770, speech-gated loudness adopts an alternative gating mechanism, focusing on speech passages. First, a \ac{VAD} is used to estimate the presence of speech in the mix~\cite{robinson2005automated}. Then, the loudness of the program is measured by considering only portions where speech is detected as active (possibly jointly with \ac{ME}).

\begin{figure}[t]
\begin{center}
\includegraphics[width=0.85\columnwidth]{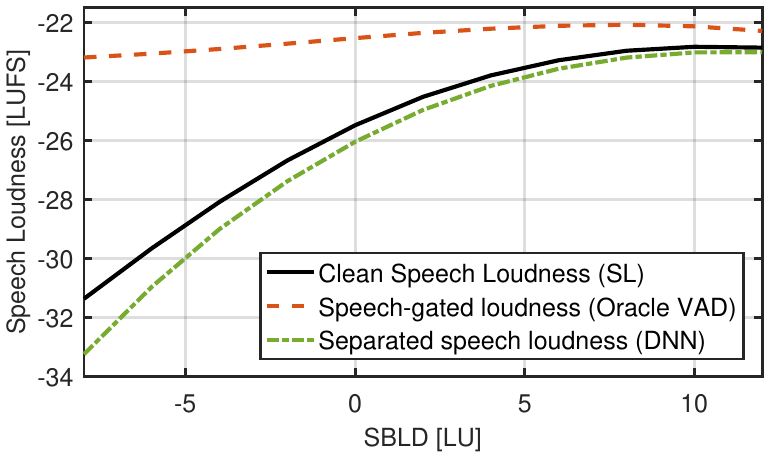}
\caption{Speech-gated loudness overestimates by nature \acf{SL} when speech is mixed together with \acf{ME}. Program loudness is $-23$\,LUFS. Oracle \acf{VAD}, i.e., the ground-truth speech activity signal is used for gating.
}
\label{fig:SGL}
\end{center}
\end{figure}

This strategy has proven useful over the years, and speech-gated loudness has become one of the most used loudness-based measures.
When speech is present without any background, speech-gated loudness equals the true loudness of speech, i.e., \ac{SL}. Speech-gated loudness is also a good approximation when speech is mixed together with background sounds with high \ac{SBLD}. It follows that over the duration of a full movie, integrated speech-gated loudness can be a reliable proxy for integrated \ac{SL}.
This is not the case when speech is locally mixed at a low relative level. 
In this case, even considering a perfect (\textit{oracle}) \ac{VAD}, deviations from \ac{SL} can be significant, as shown in \cref{fig:SGL}.
Speech-gated loudness consistently overestimates \ac{SL}, especially for \ac{SBLD}\,$<5$\,LU. This overestimation was also documented by empirical measurements on real-world movies~\cite{skovenborg2013level}.
This can be a problem if the intention is to identify local passages with low \ac{SBLD}, as we intend to do in this work.

\subsection{Estimating Clean Speech Loudness}
In the past, there were few alternatives to speech-gated loudness for cases in which speech is not available as a clean signal.
Nowadays, speech loudness can be estimated much more precisely.
For example, \cite{uhle2020clean} proposes \acp{DNN} to estimate \ac{SL} directly from the final soundtrack, i.e., without estimating the speech signal.
In this work, we propose to use a \ac{DNN} trained to separate the full clean speech signal from a final soundtrack, e.g., \cite{torcoli2021dialog}.
Consequently, the separated speech signal could be used to adjust the audio mix, if desired, or to enable personalization at the user's end via \ac{OBA}~\cite{rieger2023dialogue}.
As shown in \cref{fig:SGL}, measuring the loudness of the separated speech provides a very good estimate of \ac{SL}.

\begin{figure}[t]
\begin{center}
\includegraphics[width=0.80\columnwidth]{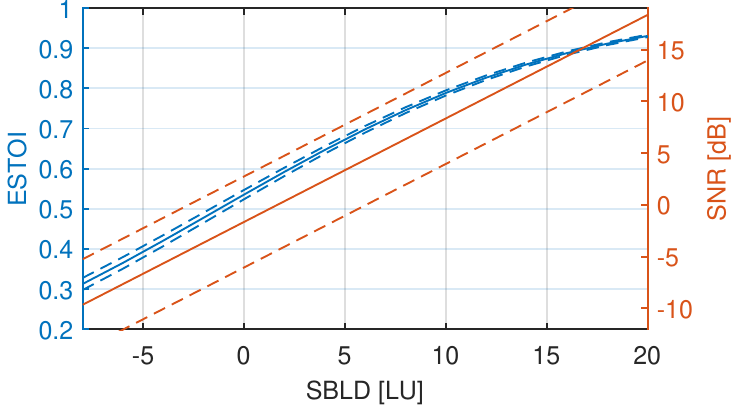}
\caption{Average (solid line) ± standard deviation
(dashed lines) showing the relation between \acf{SBLD}, \acf{ESTOI}, and \ac{SNR} for broadcast-like mixtures of speech and background signals.}
\label{fig:testa_b}
\end{center}
\end{figure}

\subsection{Loudness Range (LRA)}
Recommending a loudness value integrated over the full program cannot prevent the program or the speech signal from having local passages that are too low or too high in level.
To alleviate this problem, the \ac{LRA} is often considered~\cite{ebu_3342}. 
\ac{LRA} describes the variation or dynamics of loudness over the duration of a program. 
\ac{LRA} can be important for speech, as it can describe the distance between a loud event (e.g., an explosion or screamed dialog) and some softly whispered dialog encountered in two distinct instants of the same program. If the \ac{LRA} is too large, the audience might lower the playback volume during the loud effect, and then miss the whispered dialog.

\ac{LRA} can be compressed during playback, e.g., depending on the playback device and listening environment. \ac{LRA} control is a feature of modern audio codecs~\cite{netflixLRA, kuech2015dynamic}.
Furthermore, the program \ac{LRA} can be kept within certain ranges during the production itself, e.g., as recommended in~\cite{netflixMixSpecs}.
Controlling the \ac{LRA} can bring great advantages to the homogeneity of speech levels, especially when speech \ac{LRA} can be accurately measured. 
In cases where the clean speech signal is unavailable, the \ac{DNN}-separated speech can be used instead. Nowadays, program \ac{LRA} or speech-gated \ac{LRA} are usually employed, which can deviate significantly from the speech \ac{LRA} for the reasons discussed in \cref{sec:gating} and \cref{fig:SGL}.

\begin{figure}[t]
\begin{center}
\includegraphics[width=0.78\columnwidth]{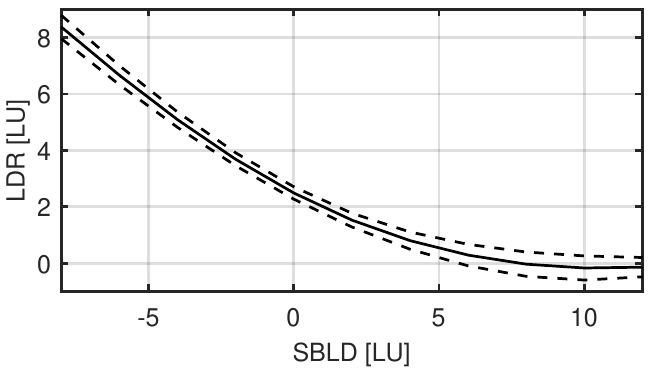}
\caption{Average \acf{LDR} $\pm$ standard deviation (dashed line) as function of \ac{SBLD}.
Both \ac{LDR} and \ac{SBLD} use integrated loudness.
\ac{LDR} settles at $0$\,LU for higher \ac{SBLD}, as clean speech loudness dominates program loudness.
}
\label{fig:SGL_b}
\end{center}
\end{figure}

\subsection{Relative Speech Loudness}
The relative level between speech and background elements in the mix has also been the focus of research and recommendations, e.g., \cite{torcoli2019preferred} and references therein.
In this work, we use the \ac{SBLD}, which is strongly related to \ac{SNR} (\cref{fig:SNR}) and to objective measures of intelligibility of speech in noise, e.g., \ac{ESTOI}~\cite{jensen2016algorithm} (\cref{fig:testa_b}).
More recently, the \ac{EBU} proposed to consider the difference between the overall program loudness and the dialog loudness, also known as \ac{LDR} \cite{ebu_s4}. 
Both \ac{SBLD} and \ac{LDR} are ways to express the relative speech loudness.
\cref{fig:SGL_b} shows the relation between \ac{LDR} and \ac{SBLD} for the data described in \cref{sec:data}.

For cinematic content, the \ac{EBU} suggests \ac{LDR} below $5$\,LU, while it is reported to sometimes reach $15$\,LU or more~\cite{ebu_s4}.
A minimum \ac{SBLD} of $4$\,LU is recommended by the \ac{DPP} in the UK and by older Netflix recommendations~\cite{torcoli2019preferred}.
It is important to note that these values are only general guidelines and not strict rules.
There may be cases and signals where the optimal \ac{SBLD} or \ac{LDR} could deviate significantly from the recommended values.
In these cases, the know-how of an experienced audio engineer is irreplaceable.
Nowadays, this human experience cannot be easily combined with the existing recommendations on \ac{SBLD} or \ac{LDR} for two reasons:
\begin{enumerate}
    \item Tools and plug-ins usually employed during production and delivery do not provide \ac{SBLD} or \ac{LDR}, as they are not trivially measured when the clean speech signal is not separately available.
    \item Single values for \ac{SBLD} or \ac{LDR} integrated over the full program do not indicate whether and where speech might be \textit{locally} critical.
\end{enumerate}
Both points are addressed by the proposed framework.

\section{Proposed Framework}
\label{sec:proposed}

\begin{figure}[t]
\centering
\includegraphics[width=1\linewidth]{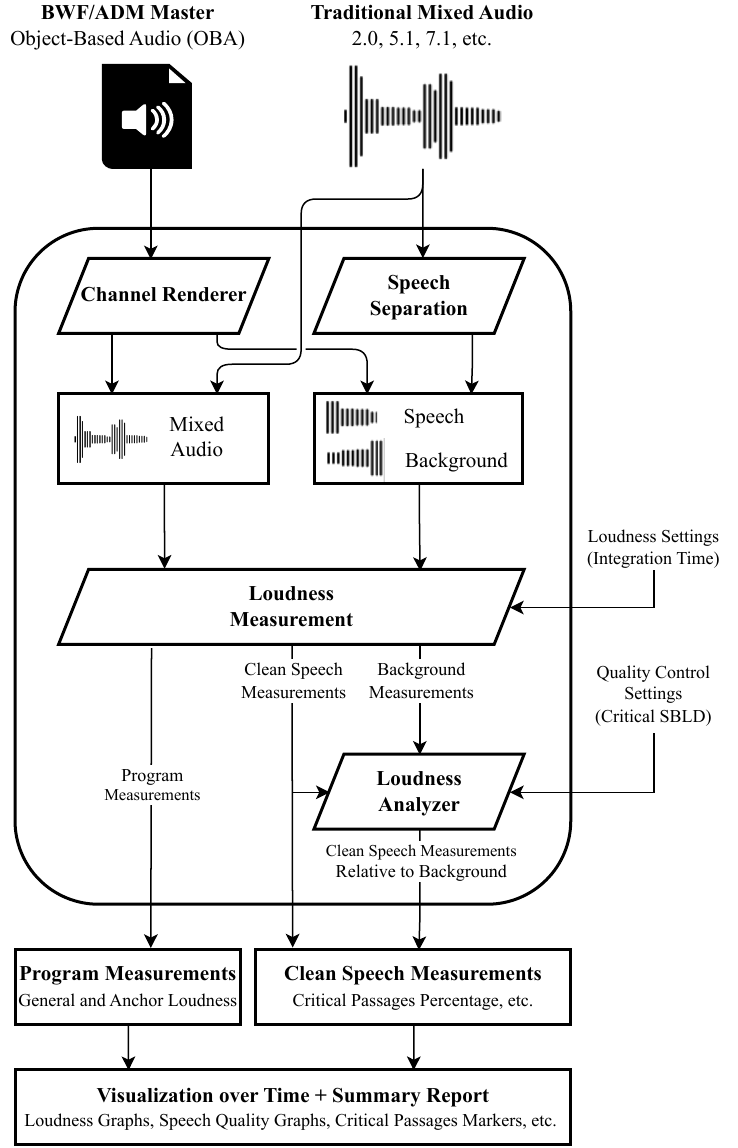}
\caption{Proposed method overview.
}
\label{fig:sys}
\end{figure}

An overview of the proposed framework is depicted in \cref{fig:sys}. The core is an accurate \ac{SL} measurement over time.
As discussed in \cref{sec:gating}, if the clean speech signal is not available from production (e.g., as \ac{OBA}), we use a \ac{DNN} trained to separate speech from a final soundtrack, e.g., \cite{torcoli2021dialog}. We consider all other sound sources as \textit{background}.
Considering the speech and background signals, we can define loudness-based macroscopic and microscopic measures.
Macroscopic measures provide a high-level summary of the full program and the speech in it, while microscopic measures are specific to time locations and are used to identify critical passages.

\subsection{Macroscopic Measures}

\begin{table*}[t]
\caption{\label{tab:macro}Macroscopic measures for loudness-based \acl{QC}.}
\begin{footnotesize}
\begin{tabular}{*{1}{l}>{\raggedright\arraybackslash}p{8.5cm}}
    \toprule
    \textbf{Quality control measure [Units]} & \textbf{Description} \\ \midrule
    \textbf{Program Measures} & \\ \midrule
    Program loudness [LUFS] & Integrated loudness of the full program~\cite{itu_bs1770}.
    \vspace{0.5em} 
    \\ 
    Program \acf{LRA} [LU] & Variation of loudness over time~\cite{ebu_3342}.
    \vspace{0.5em}
    \\ 
    True peak and sample peak [dBTP / dBFS] & Maximum true or sample peak, useful to assess the highest levels in the program.    \vspace{0.5em}
    \\ 
    Max. momentary/short-term loudness [LUFS] & Maximum local loudness values, useful for assessing the loudest events in the program. \vspace{0.5em}
    \\ 
    Speech-gated loudness [LUFS] &  Integrated loudness of the portions of programs where speech is detected.    \vspace{0.5em}
    \\ 
    Speech-gated \ac{LRA} [LU] & Variation of the speech-gated loudness.    \vspace{0.5em}
    \\ 
    \midrule
    \textbf{Clean Speech Measures} & \\ \midrule
    \Acf{SL} [LUFS] & Integrated loudness of clean speech signal (possibly separated, e.g., using \acp{DNN}).    \vspace{0.5em}
    \\ 
    Clean speech \ac{LRA} [LU] & Variation of the clean speech loudness (possibly separated, e.g., using \acp{DNN}).    \vspace{0.5em}
    \\ 
    \acf{LDR} [LU] & Program loudness minus clean speech loudness, useful to assess the speech level relative to the one of the full program \cite{ebu_s4}.    \vspace{0.5em}
    \\    
    \acf{SBLD} [LU] & \ac{SL} minus integrated background loudness, useful to assess the relative speech level in the mix, similarly to the concept of \ac{SNR}, as shown in \cref{fig:testa_b}. We compute \ac{SBLD} considering only the passages where speech is active. \ac{SBLD} is also used as a local (microscopic) measure when computed using shorter integration windows.   \vspace{0.5em}
    \\ 
    Critical speech percentage [\%] & Temporal percentage of the critical speech duration relative to the overall speech duration. Speech is considered critical when its local SBLD and SLD are below the given threshold.     \vspace{0.5em}
    \\ 
    \bottomrule
\end{tabular}
\end{footnotesize}
\end{table*}

\cref{tab:macro} gives an overview of the macroscopic measures that we consider. While the measures listed under \textit{Program Measures} are already used in the industry, the measures listed under \textit{Clean Speech Measures} are either novel or not used in practice, as they are hard to estimate. 
We propose to enable them via \ac{DNN}-based speech signal estimates in the case clean speech is not available.

\begin{figure}[t]
\centering
\includegraphics[width=1\linewidth]{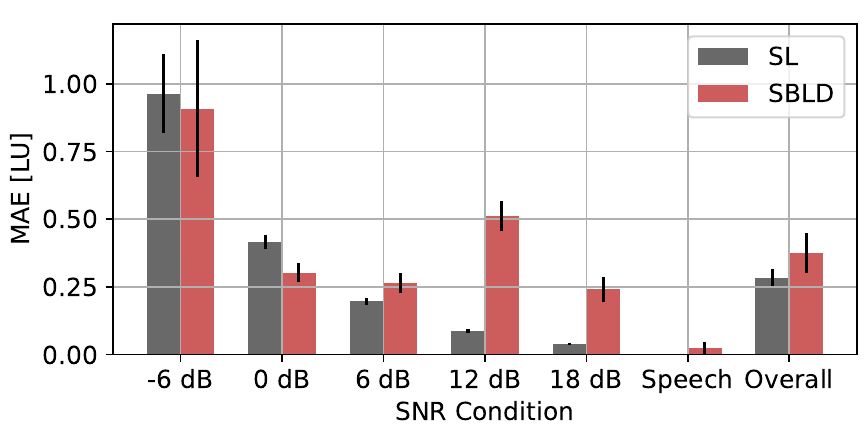}
\caption{\label{fig:MAE}\acf{MAE} and standard deviation for the DNN-based estimation of \textit{integrated} \acf{SL}, and \acf{SBLD}.
Different mixing levels of speech and background are considered along with speech only and overall, i.e., pooling all previous cases.}
\end{figure}

In particular, integrated \ac{SL} and \ac{SBLD} are fundamental measures to assess the absolute and relative level of speech over the full program.
The performance of the estimation based on the DNN-separated signals is analyzed in \cref{fig:MAE} for different \ac{SNR} conditions. The \ac{MAE} between the ground-truth \ac{SL} and \ac{SBLD} and the DNN-based estimates is considered as a performance indicator, as also done in~\cite{uhle2020clean}.
\cref{fig:MAE} shows that \ac{MAE} is below $1.0$\,LU in all cases when using DNN-separated signals instead of the clean signals from production.

As an additional macroscopic measure, we introduce the concept of critical speech percentage. This reports the temporal percentage of critical speech duration relative to the overall speech duration. Speech is considered critical if its local \ac{SLD} or its local \ac{SBLD} are below certain thresholds, as explained in \cref{sec:micro}.

\subsection{Microscopic Measures}
\label{sec:micro}

In this work, we use short-term loudness to define microscopic (or local) measures. Different integration times may also be effective and preferred.
A more detailed discussion on the possible integration times is left for future work.

First, the short-term loudness of the speech and background signals are measured.
To remove the effect of the overall program level, the integrated \ac{SL} is subtracted from the short-term speech loudness to obtain the \ac{SLD}.
It follows that the \ac{SLD} measures the absolute local level of speech on a scale where 0\,LU corresponds to the integrated clean speech loudness. Large deviations away from the 0\,LU reference point should be avoided. Higher levels could force the audience to reduce the overall level. Most importantly, lower levels indicate segments where speech is hard to hear, as its absolute level is too low.
The \ac{SLD} can be seen as the local counterpart of the speech \ac{LRA}. While the speech \ac{LRA} is a macroscopic measure summarizing the variation of speech loudness over the entire program, the \ac{SLD} shows this deviation across time and allows us to locate passages that might be problematic.

The second core measure is the local \ac{SBLD}. While the \ac{SBLD} can also be useful as a macroscopic measure (\cref{tab:macro}), the local \ac{SBLD} perfectly complements the \ac{SLD} in providing insights into the locations of potentially critical speech passages due to their low speech level, with \ac{SLD} focusing on the absolute speech level and \ac{SBLD} focusing on the level relative to the background.
We compute the local \ac{SBLD} as the difference between the short-term loudness of the speech and that of the background signal.
We use a shorter time window (e.g., $1$\,s) to compute auxiliary speech loudness values. When these fall below a certain activation threshold (e.g., $-65$\,LUFS), the passage is considered to have no speech, and the \ac{SBLD} is not computed. Moreover, we exclude portions with extremely low \ac{SBLD}, e.g., speech is not considered active if \ac{SBLD} is below $-10$\,LU.

\begin{figure}[t]
\centering
\includegraphics[width=1\linewidth]{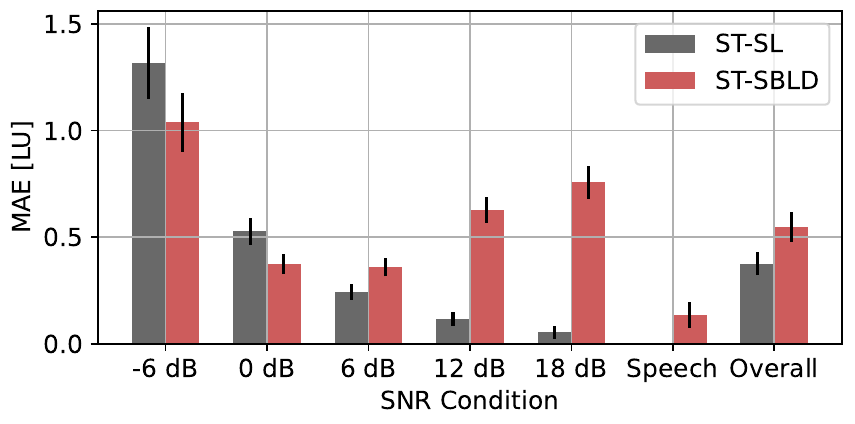}
\caption{\label{fig:MAE_ST} Short-term (ST) measures estimation.
}
\end{figure}

For the estimation of local \ac{SL} and \ac{SBLD}, we use the speech signal estimated by the \ac{DNN} if the clean speech signal is not available. The \ac{MAE} for local \ac{SL} and \ac{SBLD} is analyzed in \cref{fig:MAE_ST}, where it can be observed that for the local measures, \ac{MAE} is below $1.0$\,LU in most of the considered conditions, while it can be slightly greater than $1.0$\,LU for negative \ac{SNR} conditions.

Critical passages are defined as passages where \ac{SLD} or \ac{SBLD} are locally below given thresholds.
As a first proposal, we use $-10$\,LU as the threshold for critical \ac{SLD}, and $0$\,LU as the threshold for critical local \ac{SBLD}.
Considering real-world TV audio excerpts, \ac{SBLD} $=0$ resulted in the highest correlation between critical speech percentage and the self-reported listening effort scores gathered in \cite{torcoli2022dialogue}.
Nevertheless, we propose a flexible framework, and we leave the possibility of tuning the exact values for these thresholds and the analysis window size open. We believe that these values are better tuned by those involved in the production and \acl{QC} stages, as they can best adjust them to fit the goals and intents of the considered productions, program genre, or platforms. 

\subsection{Examples}
\begin{figure*}
\centering
\includegraphics[width=1\linewidth]{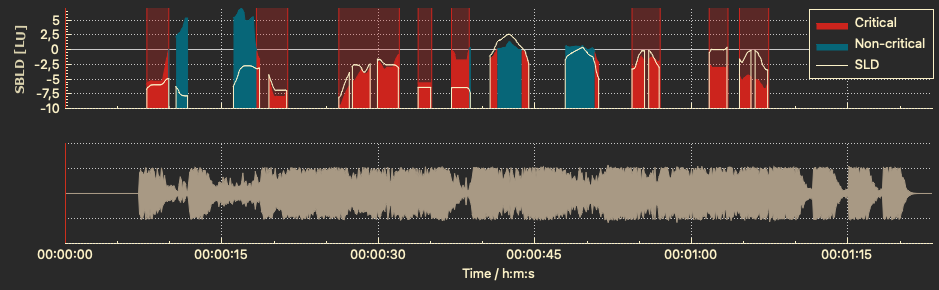}
\caption{\label{fig:inception}Trailer of \textit{Inception} by Christopher Nolan (2010, Warner Bros.): \protect\url{https://www.youtube.com/watch?v=Jvurpf91omw}. Integrated \ac{SBLD}~$=-2.9$\,LU. \ac{LDR}~$=5.6$\,LU. Critical speech percentage: $70$\%.
}
\end{figure*}

\begin{figure*}
\centering
\includegraphics[width=1\linewidth]{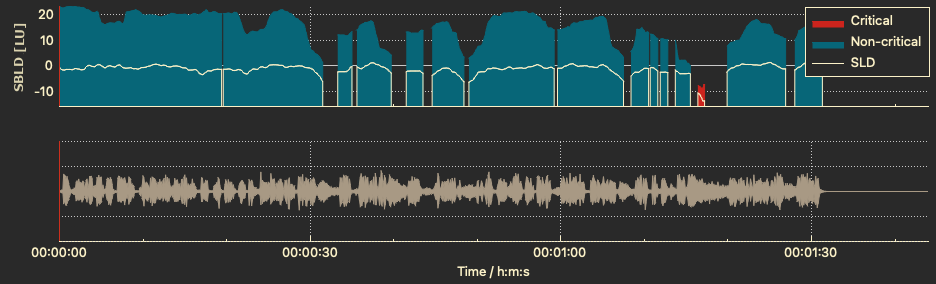}
\caption{\label{fig:ronny}Trailer of \textit{Ronny Chieng: Asian Comedian Destroys America!} by Sebastian DiNatale (2019, Netflix): \protect\url{https://www.youtube.com/watch?v=wckvy07xStI}. Integrated \ac{SBLD}~$=10.8$\,LU. \ac{LDR}~$=-0.2$\,LU. Critical speech percentage: $1$\%.
}
\end{figure*}

\begin{figure*}
\centering
\includegraphics[width=1\linewidth]{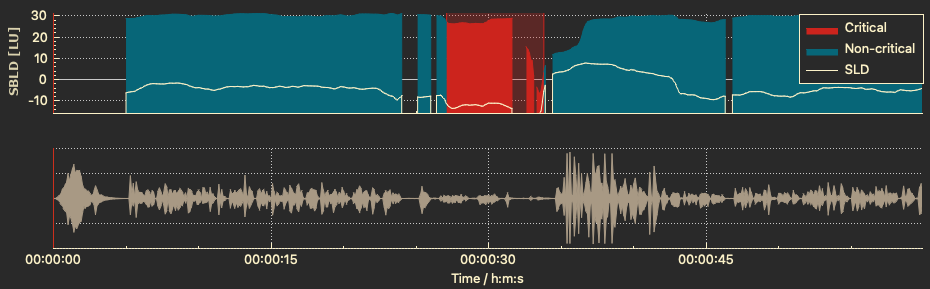}
\caption{\label{fig:learn}First minute of \textit{Learn How To Mix And Process An Extremely Dynamic Dialog Recording In Minutes} (2019, Production Expert): \protect\url{https://www.youtube.com/watch?v=P2rh-P7buhY}. Integrated \ac{SBLD}~$=13.9$\,LU. \ac{LDR}~$=0.5$\,LU. Critical speech percentage: $11$\%.
}
\end{figure*}

Using the proposed method, we analyzed three short examples available online.
The outcome of the analysis is shown in \cref{fig:inception,fig:ronny,fig:learn}. In the figure captions, links to the original content are given so that the reader can compare the audio with the corresponding analysis.

The first example is a very cinematic trailer with many sound effects, resulting in $70$\% of critical passages (\cref{fig:inception}). The second example is a trailer for a stand-up comedy show, where only one short sentence ("It's me..." at minute 1:16) is detected as critical (\cref{fig:ronny}).
Finally, we consider the first minute of a tutorial on how to mix extremely dynamic speech. The tutorial presents a speech recording that has a passage with low \ac{SLD} (and no background at all). This is correctly identified as critical by the proposed method (\cref{fig:learn}).

\section{Conclusion}
This paper proposes expanding the set of loudness-based measures, with a particular focus on speech. 
Given that clean speech is rarely available in practice, we demonstrate that it can be separated from the final soundtrack using a \ac{DNN}.
The separated speech can be used to reliably estimate \ac{SL} and \ac{SBLD}, both as integrated (macroscopic) and as local (microscopic) measures. In most of the tested conditions, the \ac{MAE} is below $1.0$\,LU.
Finally, critical passages are defined based on the local \ac{SLD} and the local \ac{SBLD}.

The provided examples qualitatively show that detecting critical passages supports evaluating and controlling the audio quality during and after content production.
This can be a powerful complement to other measures commonly used in the industry, such as program and speech loudness or \ac{LRA}.

\balance
\bibliographystyle{IEEEtran}
\bibliography{references}

\begin{thebibliography}{10}
\providecommand{\url}[1]{#1}
\csname url@samestyle\endcsname
\providecommand{\newblock}{\relax}
\providecommand{\bibinfo}[2]{#2}
\providecommand{\BIBentrySTDinterwordspacing}{\spaceskip=0pt\relax}
\providecommand{\BIBentryALTinterwordstretchfactor}{4}
\providecommand{\BIBentryALTinterwordspacing}{\spaceskip=\fontdimen2\font plus
\BIBentryALTinterwordstretchfactor\fontdimen3\font minus
  \fontdimen4\font\relax}
\providecommand{\BIBforeignlanguage}[2]{{%
\expandafter\ifx\csname l@#1\endcsname\relax
\typeout{** WARNING: IEEEtran.bst: No hyphenation pattern has been}%
\typeout{** loaded for the language `#1'. Using the pattern for}%
\typeout{** the default language instead.}%
\else
\language=\csname l@#1\endcsname
\fi
#2}}
\providecommand{\BIBdecl}{\relax}
\BIBdecl

\bibitem{torcoli2021dialog}
M.~Torcoli, C.~Simon, J.~Paulus \emph{et~al.}, ``Dialog+ in broadcasting: First
  field tests using deep-learning-based dialogue enhancement,'' in
  \emph{International Broadcasting Convention (IBC) Technical Papers}, 2021.

\bibitem{mapp2016intelligibility}
P.~Mapp, ``Intelligibility of cinema \& tv sound dialogue,'' in \emph{Audio
  Engineering Society Convention 141}, 2016.

\bibitem{mathers1991study}
C.~D. Mathers, ``A study of sound balances for the hard of hearing,'' in
  \emph{BBC White Paper, Report 1991-03}, 1991.

\bibitem{thornton:loudness}
M.~Thornton, ``Loudness - everything you need to know,'' May 2021,
  \url{https://www.production-expert.com/production-expert-1/loudness-everything-you-need-to-know}
  [Accessed: Feb. 2024].

\bibitem{ward2019casualty}
L.~Ward, M.~Paradis, B.~Shirley, L.~Russon, R.~Moore, and R.~Davies, ``Casualty
  accessible and enhanced {(A\&E)} audio: Trialling object-based accessible tv
  audio,'' in \emph{Audio Engineering Society Convention 147}, 2019.

\bibitem{rieger2023dialogue}
D.~Rieger, C.~Simon, M.~Torcoli, and H.~Fuchs, ``Dialogue enhancement with
  {MPEG-H A}udio: An update on technology and adoption,'' in \emph{Audio
  Engineering Society Convention 154}, 2023.

\bibitem{amazon2023}
{About Amazon}, ``Prime video launches a new accessibility feature that makes
  it easier to hear dialogue in your favorite movies and series,'' April 2023,
  \url{https://www.aboutamazon.com/news/entertainment/prime-video-dialogue-boost}
  [Accessed: Feb. 2024].

\bibitem{ebu_s4}
{EBU~R~128~S4}, ``Loudness normalisation of cinematic content,'' European
  Broadcasting Union ({EBU}), Nov. 2023.

\bibitem{itu_bs1770}
{ITU-R}, ``Recommendation {ITU-R BS.}1770-5: Algorithms to measure audio
  programme loudness and true-peak audio level,'' Int. Telecommunication Union
  (ITU), Radiocommunication Sector, Nov. 2023.

\bibitem{loizou2007}
P.~C. Loizou, \emph{Speech Enhancement: Theory and Practice}.\hskip 1em plus
  0.5em minus 0.4em\relax CRC press, 2007.

\bibitem{ohlenforst2018impact}
B.~Ohlenforst, D.~Wendt, S.~E. Kramer \emph{et~al.}, ``Impact of {SNR}, masker
  type and noise reduction processing on sentence recognition performance and
  listening effort as indicated by the pupil dilation response,'' \emph{Hearing
  research}, vol. 365, pp. 90--99, 2018.

\bibitem{klink2012measuring}
K.~B. Klink, M.~Schulte, and M.~Meis, ``{Measuring listening effort in the
  field of audiology — A literature review of methods (part 1)},''
  \emph{Z.~Audiol.}, vol.~51, no.~2, pp. 60--67, 2012.

\bibitem{nilsson1994}
M.~Nilsson, S.~D. Soli, and J.~A. Sullivan, ``Development of the hearing in
  noise test for the measurement of speech reception thresholds in quiet and in
  noise,'' \emph{The Journal of the Acoustical Society of America}, vol.~95,
  no.~2, pp. 1085--1099, 1994.

\bibitem{alhanbali2019measures}
S.~Alhanbali, P.~Dawes, R.~E. Millman, and K.~J. Munro, ``Measures of listening
  effort are multidimensional,'' \emph{Ear and Hearing}, vol.~40, no.~5, pp.
  1084--1097, 2019.

\bibitem{torcoli2022dialogue}
M.~Torcoli, T.~Robotham, and E.~A.~P. Habets, ``Dialogue enhancement and
  listening effort in broadcast audio: A multimodal evaluation,'' in \emph{2022
  14th International Conference on Quality of Multimedia Experience (QoMEX)},
  2022.

\bibitem{rood2006flight}
G.~M. Rood and S.~H. James, ``In-flight communication,'' \emph{Ernsting's
  Aviation Medicine, fourth edition. Edward Arnold (Publishers) Ltd, London
  (UK)}, pp. 385--394, 2006.

\bibitem{ebu_r128}
{EBU~R~128}, ``Loudness normalization and permitted maximum level of audio
  signals,'' European Broadcasting Union ({EBU}), Nov. 2023.

\bibitem{berendes2022validating}
H.~U. Berendes, A.~Travaglini, and C.~Uhle, ``Validating loudness alignment via
  subjective preference: Towards improving itu-r bs. 1770-4,'' in \emph{Audio
  Engineering Society Convention 153}, 2022.

\bibitem{atsc_a85}
{ATSC~Doc~A/85}, ``{ATSC R}ecommended practice: Techniques for establishing and
  maintaining audio loudness for digital television,'' Advanced Television
  Systems Committee ({ATSC}), Mar. 2013.

\bibitem{cramer2010}
F.~Camerer, ``On the way to loudness {N}irvana - {A}udio levelling with {EBU R
  128},'' in \emph{EBU Technical Review - Q3}, 2010.

\bibitem{ebu_3341}
{EBU~Tech~Doc~3341}, ``Loudness metering: {'EBU Mode'} metering to supplement
  loudness normalisation in accordance with {EBU R 128},'' European
  Broadcasting Union ({EBU}), Nov. 2023.

\bibitem{aes77_2023}
{AES77-2023}, ``{AES} recommended practice loudness guidelines for internet
  audio streaming and on-demand distribution,'' Audio Engineering Society
  {(AES)}, Jul. 2023.

\bibitem{riedmiller2005practical}
J.~Riedmiller, C.~Robinson, A.~Seefeldt, and M.~Vinton, ``Practical program
  loudness measurement for effective loudness control,'' in \emph{Audio
  Engineering Society Convention 118}, 2005.

\bibitem{aes71_2018}
{AES71-2018}, ``{AES} recommended practice loudness guidelines for over the top
  television and online video distribution,'' Audio Engineering Society
  {(AES)}, Jul. 2018.

\bibitem{netflixMixSpecs}
``Netflix sound mix specifications \& best practices v1.5,'' Netflix Partner
  Help Center, Nov. 2023,
  \url{https://partnerhelp.netflixstudios.com/hc/en-us/articles/360001794307-Netflix-Sound-Mix-Specifications-Best-Practices-v1-5}
  [Accessed: Feb. 2024].

\bibitem{netflixLRA}
P.~Williams and V.~Gondi, ``Optimizing the aural experience on android devices
  with {xHE-AAC},'' Netflix Technology Blog, Jan. 2021,
  \url{https://netflixtechblog.com/optimizing-the-aural-experience-on-android-devices-with-xhe-aac-c27714292a33}
  [Accessed: Feb. 2024].

\bibitem{apple:anchor}
{Apple Developer Documentation}, ``Adjusting anchor loudness,'' --,
  \url{https://developer.apple.com/documentation/http-live-streaming/adjusting-anchor-loudness}
  [Accessed: Feb. 2024].

\bibitem{robinson2005automated}
C.~Robinson and M.~Vinton, ``Automated speech/other discrimination for loudness
  monitoring,'' in \emph{Audio Engineering Society Convention 118}, 2005.

\bibitem{skovenborg2013level}
E.~Skovenborg and T.~Lund, ``Level-normalization of feature films using
  loudness vs speech,'' in \emph{Audio Engineering Society Convention 135},
  2013.

\bibitem{uhle2020clean}
C.~Uhle, M.~Kratschmer, A.~Travaglini, and B.~Neugebauer, ``Clean dialogue
  loudness measurements based on deep neural networks,'' in \emph{Audio
  Engineering Society Convention 150}, 2021.

\bibitem{ebu_3342}
{EBU~Tech~Doc~3342}, ``Loudness range: A measure to supplement {EBU R 128}
  loudness normalization,'' European Broadcasting Union ({EBU}), Nov. 2023.

\bibitem{kuech2015dynamic}
F.~Kuech, M.~Kratschmer, B.~Neugebauer, M.~Meier, and F.~Baumgarte, ``Dynamic
  range and loudness control in mpeg-h 3d audio,'' in \emph{Audio Engineering
  Society Convention 139}, 2015.

\bibitem{torcoli2019preferred}
M.~Torcoli, A.~Freke-Morin, J.~Paulus \emph{et~al.}, ``Preferred levels for
  background ducking to produce esthetically pleasing audio for tv with clear
  speech,'' \emph{Journal of the Audio Engineering Society}, vol.~67, no.~12,
  pp. 1003--1011, 2019.

\bibitem{jensen2016algorithm}
J.~Jensen and C.~H. Taal, ``An algorithm for predicting the intelligibility of
  speech masked by modulated noise maskers,'' \emph{IEEE/ACM Transactions on
  Audio, Speech, and Language Processing}, vol.~24, no.~11, pp. 2009--2022,
  2016.

\end{thebibliography}

\end{document}